\documentclass[12pt]{elsarticle}
\usepackage{amsfonts,enumerate,amsmath,amssymb}
\usepackage{color,latexsym,amsfonts,amssymb}
\usepackage{ulem}
\usepackage{appendix}
\usepackage{ragged2e}

\newcommand{\bed}{\begin{displaymath}}
\newcommand{\eed}{\end{displaymath}}
\newcommand{\bea}{\bed\begin{array}{rl}}
	\newcommand{\eea}{\end{array}\eed}

\newcommand{\barray}{\begin{array}{ll}}
	\newcommand{\earray}{\end{array}}
\numberwithin{equation}{section}

\begin{document}
\begin{frontmatter}	
\title{Rate-dependent bifurcation dodging in a thermoacoustic system driven by colored noise}

\author[author1]{Xiaoyu Zhang}
\ead{xiao\_yu\_zhang@yahoo.com}

\author[author1,author2]{Yong Xu\corref{mycorrespondingauthor}}
\cortext[mycorrespondingauthor]{Corresponding author}
\ead{hsux3@nwpu.edu.cn}

\author[author1]{Qi Liu}
\ead{liuqi1780280327@yahoo.com}

\author[author3]{J{\"u}rgen Kurths}
\ead{Juergen.Kurths@pik-potsdam.de}

\author[author4]{Celso Grebogi}
\ead{grebogi@abdn.ac.uk}

\address[author1]{School of Mathematics and Statistics, Northwestern Polytechnical University, Xi'an, 710072, China}

\address[author2]{MIIT Key Laboratory of Dynamics and Control of Complex Systems, Northwestern Polytechnical University, Xi’an, 710072, China}

\address[author3]{Potsdam Institute for Climate Impact Research, Potsdam 14412, Germany}

\address[author4]{Institute for Complex Systems and Mathematical Biology, School of Natural and Computing Sciences, King’s College, University of Aberdeen, Aberdeen AB24 3UE, United Kingdom}

\setlength{\baselineskip}{0.26in}	
\begin{abstract}
Tipping in multistable systems occurs usually by varying the input slightly, resulting in the output switching to an often unsatisfactory state. This phenomenon is manifested in thermoacoustic systems. This thermoacoustic instability may lead to the disintegration of rocket engines, gas turbines and aeroengines, so it is necessary to design control measures for its suppression. It was speculated that such unwanted instability states may be dodged by changing quickly enough the bifurcation parameters. Thus, in this work, based on a fundamental mathematical model of thermoacoustic systems driven by colored noise, the corresponding Fokker-Planck-Kolmogorov equation of the amplitude is derived by using a stochastic averaging method. A transient dynamical behavior is identified through a probability density analysis. We find that the rate of change of parameters and the correlation time of the noise are helpful to dodge thermoacoustic instability, while a relatively large noise intensity is a disadvantageous factor. In addition, power-law relationships between the maximum amplitude and the noise parameters are explored, and the probability of successfully dodging a thermoacoustic instability is calculated. These results serve as a guide to the design of engines and to propose an effective control strategy, which is of great significance to aerospace-related fields.
\end{abstract}
\end{frontmatter}

\setlength{\baselineskip}{0.26in}
\section{Introduction}
\label{sec1}
In a multistable system, tipping usually refers to the event that a small change in input results in a sudden and disproportionate change in output \cite{ashwin2017parameter, lenton2020tipping}. Tipping was first introduced into sociology in the 1950s to describe the phenomenon of ``white escape" in American urban communities, i.e., when the number of non-white residents in a community reaches a certain level, that quickly leads to the result in which the community is completely occupied by non-white people \cite{Grodzins1957metropolitan}. After that, the publication of the book \cite{Gladwell2000tipping} made the concept of ``tipping'' be generally accepted. It is currently widely used in climatology, ecology and several other research fields \cite{ashwin2019extreme, holland2006future, clark2013light, yan2010diagnosis}. In general, one of the equilibrium points in multistable systems is desirable, and the shift to another equilibrium point leads to disastrous consequences, often irreversible to some extent, which requires control strategies to avoid or slow down the system from drifting towards a tipping \cite{li2020transition, zhang2019random, liu2020bistability}.

In view of the catastrophic consequences of tipping, the control of tipping has always been a timely research topic in various fields. Among them, the effective means is to identify early warning signals for a timely control \cite{ma2018detecting, ma2019predicting, ma2019slowing, ma2020quantifying}. In addition, time-delay feedback control strategy is proposed to mitigate the tipping \cite{farazmand2020mitigation}. In a pollinator-plant mutualistic network, it was shown that Gaussian white noise is an important factor to promote the state recovery after tipping, so as to avoid the adverse consequences of tipping \cite{meng2020noise-enabled}. Due to the complexity of multistable systems, a single control strategy may not be suitable for all systems. Therefore, it is necessary to determine the main factors in a specific system, and then put forward effective control strategies to dodge the undesirable states.

Tipping in thermoacoustic systems was shown to result from thermoacoustic instability in combustion chambers \cite{mcmanus1993a}. When there is a positive feedback between the fluctuation of sound pressure and unsteady heat release rate in the combustion chamber, a thermoacoustic instability occurs \cite{lieuwen2012unsteady}. It may lead to a structural damage of rocket and aeroengine, compromising rocket launching missions due to engine disintegration, which is clearly an unwanted state in thermoacoustic systems. As an important example, in the Apollo program of the United States, more than 2000 full-scale engine experiments have been conducted on the F-1 engine due to the thermoacoustic instability, consuming a lot of human and material resources \cite{oefelein1993comprehensive}. Hence, the most important and significant reason to study tipping in thermoacoustic systems is to fully understand the observed thermoacoustic instability and to learn how to influence and control it.

There are two kinds of control strategies for thermoacoustic instability: passive and active control \cite{mcmanus1993a}. Passive control reduces the instability mainly by decreasing the combustion response of propellant and increasing the structural damping. The design process of passive control consumes a lot of resources and time, and the control results obtained only have effect on specific engines in a certain working range. Thus passive control is mainly applicable to a situation where the working environment is detrimental and an active control method is difficult to be enacted. The active control of thermoacoustic instability is based on a certain control algorithm, e.g. by periodically adding energy to the combustion system to suppress oscillation. It requires a mathematical model to describe the dynamical behavior of the thermoacoustic system. Through the mathematical model, the dynamic characteristics of the combustion chamber under different working conditions is obtained, and then the active control function of the thermoacoustic instability is inferred. It should be noted that via the traditional theoretical methods it is very difficult to predict the thermoacoustic instability in thermoacoustic systems, and the classical linear stability method is only suitable to study asymptotic instability. Moreover, the dynamics of the nonlinear thermoacoustic instability belongs to the transient growth process, and the coupling between the sound field and the transient burning rate of propellant needs to be considered, which makes the analysis even more difficult. The transient characteristics and the nonlinear behaviors are very important and fundamental for understanding the thermoacoustic instability, which is the reason why we mainly study the transient dynamical behaviors of the nonlinear mathematical model of the thermoacoustic system.

When the combustion chamber produces a nonlinear thermoacoustic instability, except for a few cases which leads to engine damage, the amplitude  generally stays at a finite value, and the system has a bounded motion, usually showing  periodic limit cycle oscillation. For practical applications, it is desirable to understand the magnitude of this finite amplitude motion and how it is affected by system parameters. This helps to actively change the parameters to reduce the amplitude of the oscillation, so as to dodge the thermoacoustic instability. Considering the non-autonomous dynamical characteristics of the thermoacoustic system, the rate of change of the parameter is likely to be an important factor affecting the amplitude of the thermoacoustic instability. 

A variety of rate-dependent bifurcations are available in the literature. From the perspective of mathematical theory, Ashwin et al. \cite{ashwin2017parameter}, based on the concept of a pull-back attractor, gave the mathematical definition of rate-induced tipping for a class of nonautonomous systems. Kiers \cite{kiers2019rate-induced} put forward the definition for discrete dynamical systems, and gave the condition of rate-induced tipping. Sujith \cite{manikandan2020rate} showed experimental evidence of rate-dependent tipping in a typical turbulent afterburner model. 
In addition, through numerical simulation, Vanselow et al. \cite{vanselow2019when} confirmed that the  Rosenzweig-MacArthur predator-prey model experienced rate-dependent bifurcation, which led to the collapse of predator and prey populations. Suchithra et al. \cite{Suchithra2020rate} studied the rate-dependent tipping in power systems, and found that the tipping is dependent on the initial conditions of the system. Those studies on the aforesaid dynamics systems suggest that it is imperative to introduce the rate-dependent  bifurcation in thermoacoustic systems.

Crucially, it is not enough to consider only the dynamical behavior of the deterministic thermoacoustic systems. A large number of experiments clarified that the amplitude and phase of the limit cycle oscillation vary between cycles, and the parameters of the stability boundary in the combustion chamber also vary \cite{noiray2013deterministic}, which shows that the thermoacoustic system is stochastic. Although the deterministic system can also exhibit as a seemingly random motion when strange attractors exist, Curlick et al. \cite{Culick1992Combustion} used a fractional dimension test to reject the possibility that the thermoacoustic system is a chaotic system. Poinsot et al. \cite{Leyer1986experimental} carried out an experimental study on the dump combustor, and they confirmed the existence of noise in the combustor through the spatial diagram of the coherence function. Li et al. \cite{li2018effects} systematically studied the influence of background noise on the nonlinear dynamics of a thermoacoustic system with a subcritical Hopf bifurcation. All of these show that the thermoacoustic system with noise excitation is more suitable to describe its response.

Recently, Bonciolini and Noiray \cite{bonciolini2019bifurcation} considered a thermoacoustic system excited by Gaussian white noise. The system has a non-monotonic dynamical behavior, which at low and high values of the bifurcation parameter range, it has a low amplitude, and at intermediate values, it shows a high amplitude. That is to say, when the bifurcation parameter value changes from minimum to maximum value, the amplitude changes from the low state to the high state and then goes back to the low one. In the thermoacoustic system, the high amplitude of the system is closely related to the thermoacoustic instability, which is not ideal. Bonciolini and Noiray \cite{bonciolini2019bifurcation}, through experiments and numerical simulations, found that when the rate of change of the parameter is large enough, it can dodge the unwanted state in the intermediate parameter region. However, the power of the white noise with zero correlation considered in that study is infinite. But since there is no noise with infinite power, the actual noise must be ``colored". This makes it necessary to consider the influence of colored noise with non-zero correlation time on the dynamical behavior of the system.

It is essential for a complete stochastic theory to study the influence of the correlation time of noises on the properties of stochastic systems. In particular, under certain nonlinear conditions, the noise with a long correlation time makes the stochastic system behave completely different from that under white noise only \cite{mei2020steady}. But, for simplicity, when the macro variables and noises can be clearly divided into two time scales and the macro variables change slowly, the disturbance can be considered as white noise \cite{haunggi2007colored}. However, when the system can not be distinguished by the two time scales clearly, it is necessary to include colored noise. In a system with rate-dependent bifurcation, increasing the rate of parameter change reduces the time scale of the macro variables and decrease the discrimination between both time scales. For our system, colored noise is more suitable to describe the random force. This is the necessity of considering colored noise in the rate-dependent systems.

The objective of this paper is to investigate whether the rate of change of parameters is  helpful to dodge thermoacoustic instability in thermoacoustic systems excited by colored noises. For a typical thermoacoustic system model, the reduced dimension Fokker-Planck-Kolmogorov (FPK) equation is obtained by using the stochastic averaging method. Through the evolution of the probability density of the amplitude, the effect of the rate-dependent of parameters on dodging thermoacoustic instability is analyzed. In addition, we also study the influence of the correlation time and intensity of noises on rate-dependent bifurcation, and calculate the probability of successfully dodging thermoacoustic instability.

This paper is organized as follows. In Section 2, thermoacoustic instability and a classical mathematical model of thermoacoustic systems are introduced. The influence of the parameter varying rate on dodging thermoacoustic instability is examined in Section 3. Section 4 discusses the dodge probability of the rate-dependent bifurcation. Finally, several conclusions are given to close this paper in Section 5.

\section{Mathematical model of the thermoacoustic system}
\label{sec2}
A thermoacoustic system is closely related to many combustion power devices, such as liquid rocket engines, solid rocket engines, gas turbines and aeroengines. These combustion power units convert chemical energy stored in the molecular bonds into kinetic energy in the engines. The first stage of energy conversion is the combustion of oxidant and fuel in the combustion chamber. At this time, the chemical energy is converted into heat energy. The second stage is realized in the nozzle. The heat energy released by combustion is converted into kinetic energy at the engine through the expansion and acceleration of the nozzle. In the first stage described above, thermoacoustic instability is where it occurs.

High performance is usually achieved by increasing the energy release rate per unit volume in the engine, where the energy density is very large. Such a large energy density may be accompanied by relatively small fluctuations, and the amplitude of these small fluctuations may be only a small disturbance. It may, however, be shown as an unacceptable large amplitude disturbance, that is, a strong pressure oscillation. Pressure oscillations are the result of the interaction of acoustic vibrations, internal combustions and flows, which are usually called thermoacoustic instability. Thermoacoustic instability can lead to abnormal interior ballistics, structural damage, engine disintegration, resulting in mission failure. This requires us to predict the thermoacoustic instability as early and as accurately as possible, and to find measures to suppress the instability without affecting the overall performance.

Fundamentally speaking, the thermoacoustic instability is essentially the coupling of the unsteady combustion and the structural acoustic characteristics of the combustion chamber. As the simplest example of thermoacoustic instability, Rijke invented and first studied the Rijke tube in 1859. Then Rayleigh proposed the Rayleigh criterion, which is the most original and scientific description of a thermoacoustic coupling effect \cite{rayleigh1878explanation}. With the development of research, the Helmholtz equation (\ref{H eq}) is often  used to describe the dynamical behavior of thermoacoustic systems in recent years,
\begin{equation}\label{H eq}
	\frac{\partial^{2} x}{\partial t^{2}}-c^{2} \nabla^{2} x=(\gamma-1) \frac{\partial q}{\partial t},
\end{equation}
where $x$, $c$, $\gamma$ and $q$ denote the acoustic pressure, the speed of sound, the heat capacity ratio and the fluctuating component of the heat release rate, respectively. After Laplace transform, orthogonal basis projection and truncated Taylor expansion approximation of the nonlinear term \cite{bonciolini2018experiments}, we get the following reduced mathematical model of the thermoacoustic system
\begin{equation} \label{main eq}
\ddot{x}-\left(v+\beta_{1} x^{2}-\beta_{2} x^{4}\right) \dot{x}+\omega_{0}^{2} x+\beta_{0} x^{3}=\xi(t),
\end{equation}
in which $v$ is bound up with the linear growth rate, and with $\beta_0$, $\beta_1$, and $\beta_2$ being real parameters. The $\xi(t)$ is the exponential correlated Gaussian colored noise with zero mean, defined as,
\begin{equation*}
\begin{aligned}
\langle\xi(t)\rangle&= 0, \\
\langle\xi(t) \xi(s)\rangle&=\frac{D}{\tau} \exp \left(-\frac{|t_1-t_2|}{\tau}\right),
\end{aligned}
\end{equation*}
where $D$ and $\tau$ are the two most important parameters for characterizing noise: noise intensity and the correlation time of the noise. 

\begin{figure}[htbp]
	\centering
	\includegraphics[width=14cm]{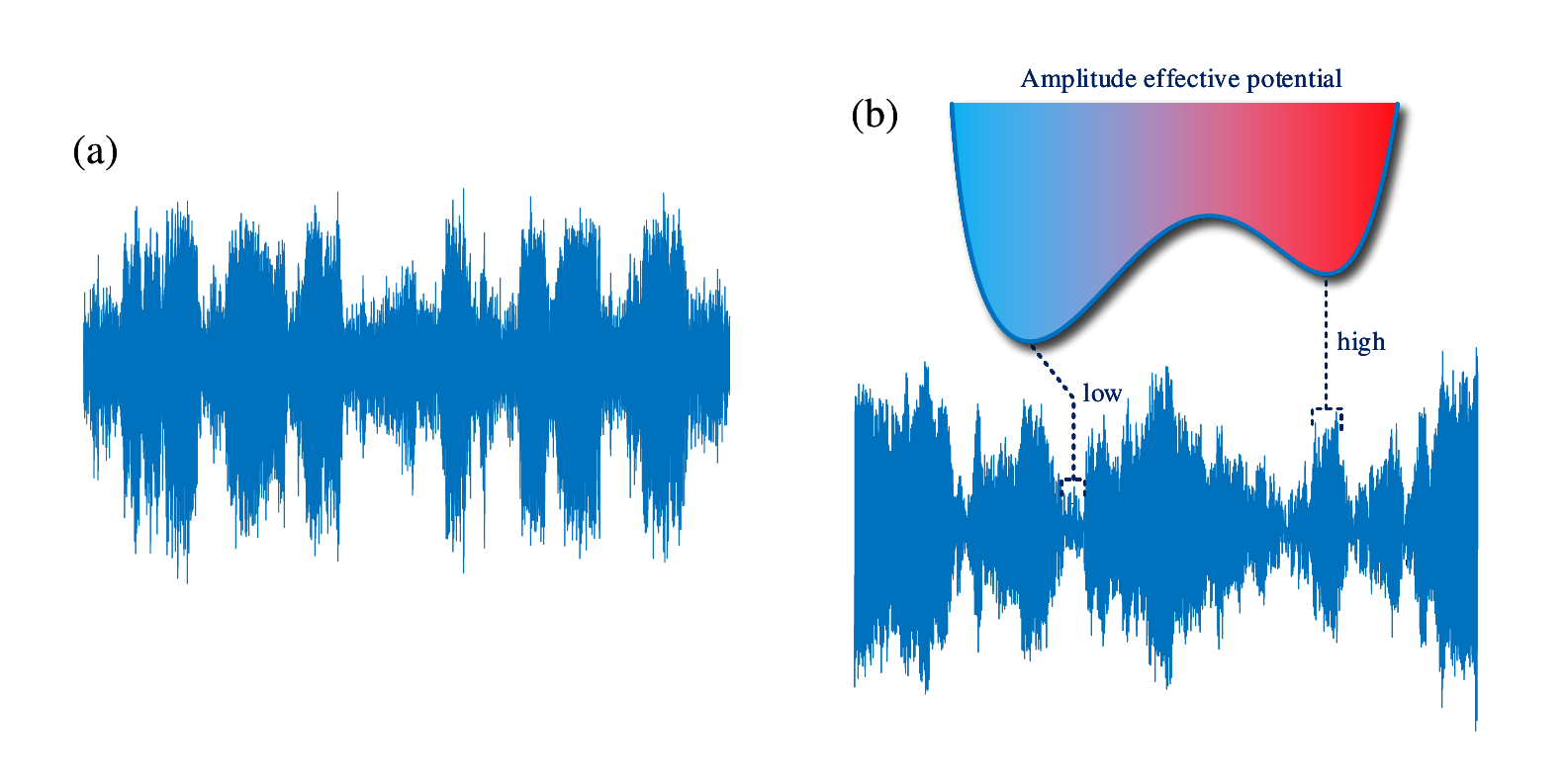}
	\caption{Time series diagram of a thermoacoustic system. (a) The experimental data; (b) The data from the mathematical model (\ref{main eq}). The two states of high and low amplitude correspond to the double well of effective potential.}
	\label{fig1}
\end{figure}

By means of the stochastic averaging method, it yields the stochastic differential equation for the amplitude $A$ of the system. Then we can derive the FPK equation (\ref{FPK}) for the amplitude $A$ with the help of stochastic dynamical theory,
\begin{equation}\label{FPK}
\frac{\partial P(A,t)}{\partial t}=-\frac{\partial}{\partial A}\left[m(A) P(A,t)\right]+\frac{1}{2} \cdot \frac{\partial^{2}}{\partial A^{2}}\left[b(A) P(A,t)\right], 
\end{equation}
with the drift and diffusion coefficients 
\begin{equation*}
\begin{aligned}
m(A)=&\frac{v}{2} A+\frac{\beta_{1}}{8} A^{3}-\frac{\beta_{2}}{16} A^{5}+\frac{D}{2 \omega_{0}^{2} A\left(1+\omega_{0}^{2} \tau^{2}\right)},\\
b(A)=&\frac{D}{\omega_{0}^{2}\left(1+\omega_{0}^{2} \tau^{2}\right)},
\end{aligned}
\end{equation*}
where $P(A,t)$ is the probability density function (PDF) of the amplitude $A$ and the $\omega_0$ is the natural angular frequency. Later, we will use the PDF $P(A,t)$ to describe the random transient dynamical behavior of the thermoacoustic system.
The amplitude effective potential function of the model (\ref{main eq}) is a double well, which corresponds to the two states of high and low amplitudes in the dynamics, as seen in the time series (see Figure \ref{fig1}(b)). This shows that the model can well describe the intermittent switching between the two states observed in the experimental thermoacoustic system \cite{bonciolini2018experiments} (Figure \ref{fig1}(a)), and further explains the rationality of the mathematical model (\ref{main eq}) used in this paper.

We consider in this work the suppression of the thermoacoustic instability of a thermoacoustic system when the linear growth rate $v$ is no longer a constant but a function of time. Through the experiments of thermoacoustic systems and the parameter identification of the results, an approximate relationship between the linear growth rate $v$ and the air mass flow rate $\dot{m}_{air}$ is obtained \cite{bonciolini2019bifurcation},
\begin{equation}\label{cos}
	v(\dot{m}_{air})=n_1 cos(n_2\dot{m}_{air})+n_3,
\end{equation}
where $n_1,n_2,$ and $n_3$ are constant coefficients. In order to be close to real situations, the $\dot{m}_{air}(t)$ is assumed to be a piecewise linear function involving a ramp 
\begin{equation}\label{linear}
	\dot{m}_{air}(t)=\dot{m}_{0}+Rt.
\end{equation} 
$R$ is the ramp rate and the $\dot{m}_{0}$ is the initial value. In the following sections, the time-varying parameter $v(t)$ is the combination of the cosine function (\ref{cos}) and the linear function (\ref{linear}).

\section{Rate-dependent bifurcation dodge with the colored noise}
\label{sec3}
In this section, we initially discuss the influence of parameter change rate $R$ on avoiding the thermoacoustic instability in a thermoacoustic system under the excitation of colored noise. The system (\ref{main eq}) can express its transient dynamical behavior through Monte Carlo simulations, but it needs a lot of calculation to collect the data. Here we use instead the reduced dimension FPK equation (\ref{FPK}) and the more efficient Crank-Nicolson difference method whose accuracy was validated in \cite{zhang2020rate-dependent}, to get the evolution of the non-autonomous thermoacoustic system. 

\begin{figure}[htbp]
	\centering
	\includegraphics[width=13cm]{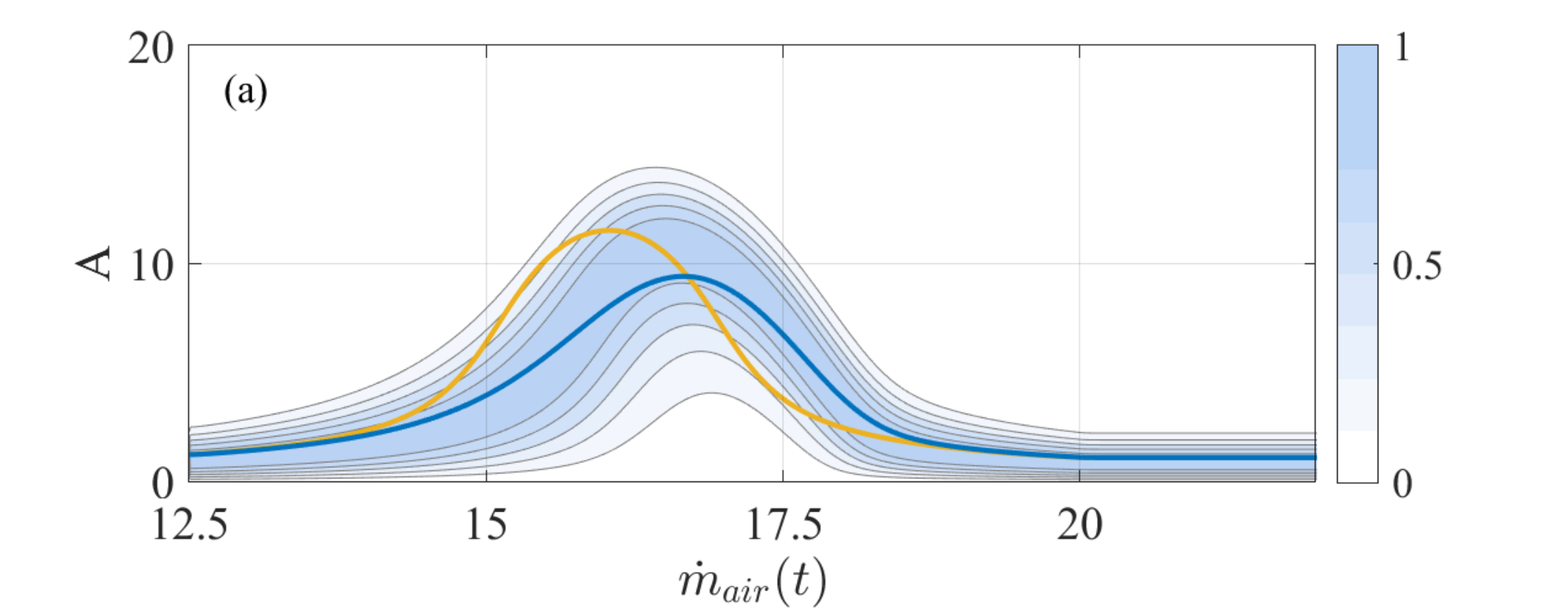}
	\includegraphics[width=13cm]{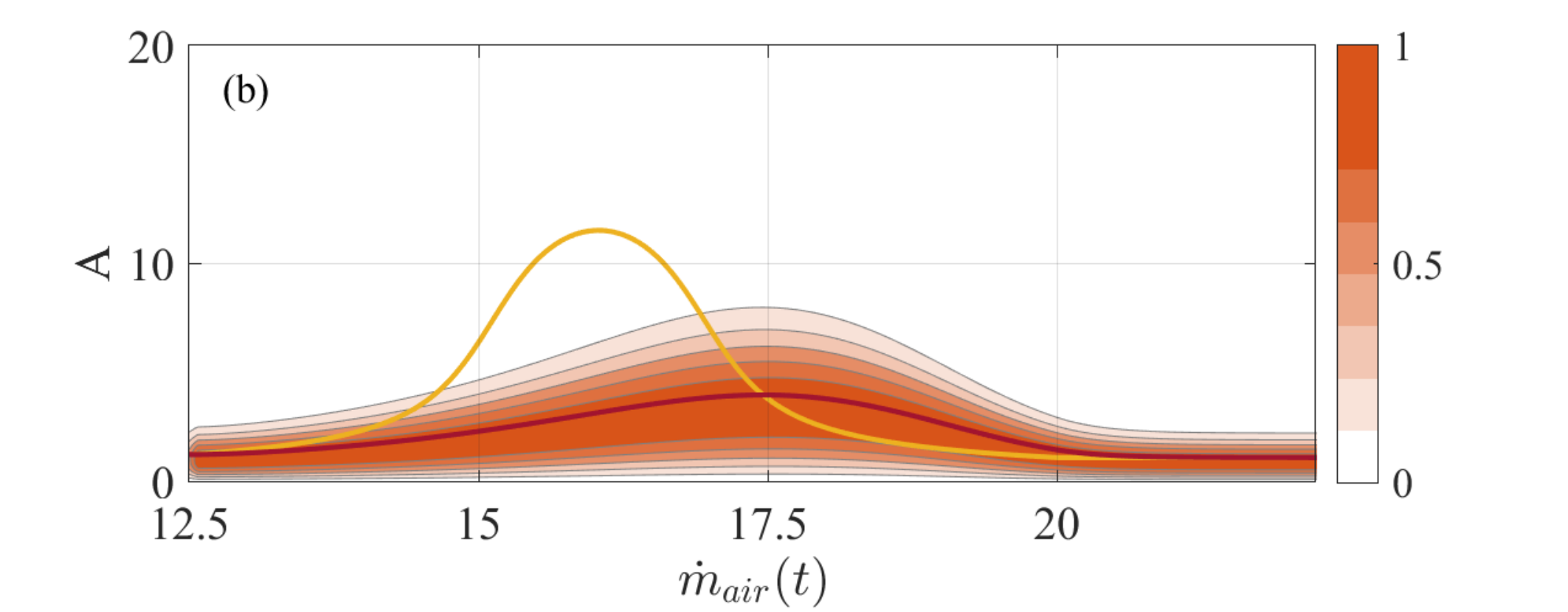}
	\caption{The effect of different rates $R$ of the time-varying parameter on the transient dynamical behavior. The contour map shows the transient probability density of the amplitude $A$ with respect to the air mass flow rate $\dot{m}_{air}$, and the value is divided by the maximum probability to make the result normalized. The yellow line is the quasi-steady deterministic results, used as a reference. Dark blue and dark red lines represent the mean value of the amplitude $A$ under different $R$. (a) $R=10$; (b) $R=75$. Other parameters are $\beta_1=-0.47$, $\beta_2=0.001$, $\omega_{0}=100\times 2\pi$, $D=8.75\times10^6$, and $\tau=0.001$. }
	\label{fig2}
\end{figure}

Figure \ref{fig2} displays the effect of different rates $R$ on the transient dynamical behavior. The time-varying parameter $v(t)$ is the combination of cosine and linear function mentioned above. The contour map shows the probability density of the amplitude $A$ changing with $\dot{m}_{air}$. The probability of dark color is large, while that of light color is small. Here we divide the probability density by its maximum value to normalize it, so its range of variation is 0 to 1. As a reference, the yellow lines, which are the same in Figures \ref{fig2}(a) and (b), are for the cases when rates are not included. Figure \ref{fig2}(a) is the probability density of the amplitude after the introduction of a relatively small rate. The dark blue line in the middle is the mean value. We find that after the introduction of the ramp rate, the amplitude is in fact reduced, and the position of the highest amplitude point is delayed. This phenomenon of rate-dependent tipping-delay has been discussed in detail in \cite{zhang2020rate-dependent}. When the rate is relatively large (see Figure \ref{fig2}(b)), it can be seen from the dark red line representing the mean value that the amplitude reduction is stronger, and the phenomenon of bifurcation dodge is more obvious.
This shows that in a non-autonomous thermoacoustic system, the thermoacoustic instability can be avoided by properly controlling the changing rate $R$ of the parameters. 

\begin{figure}[htbp]
	\centering
	\includegraphics[width=13cm]{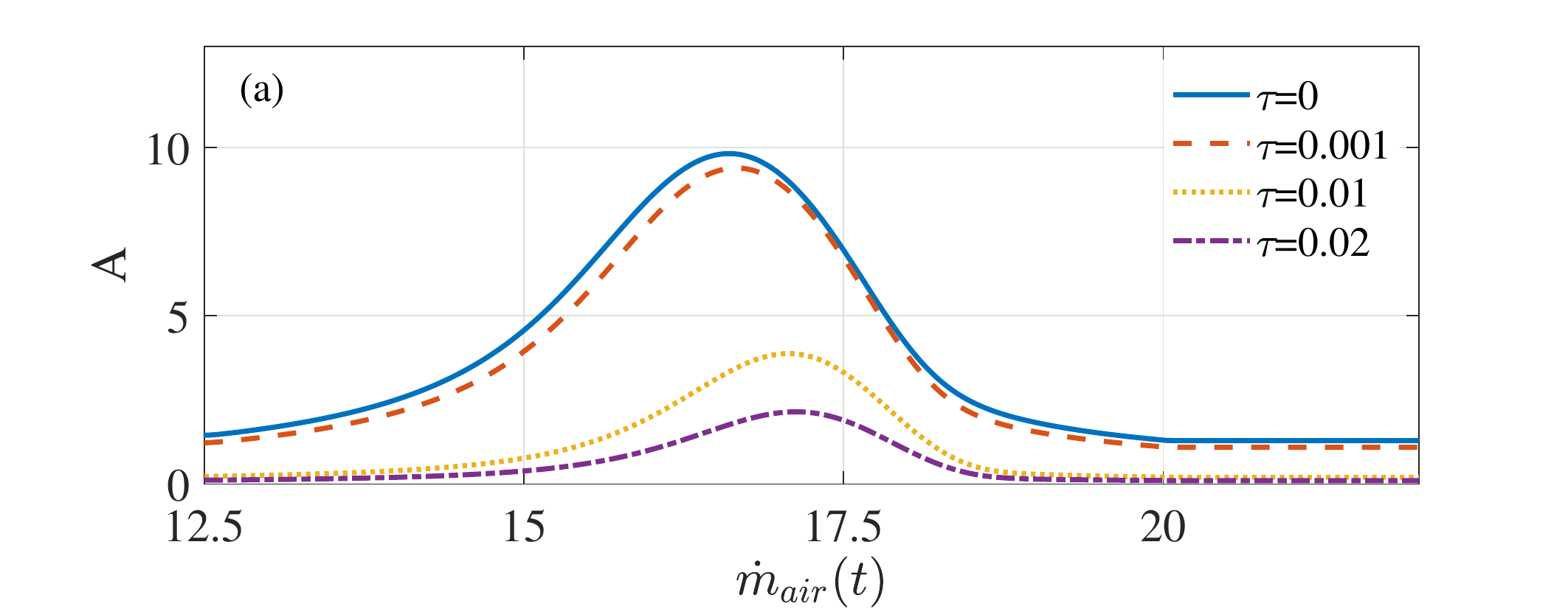}	
	\includegraphics[width=13cm]{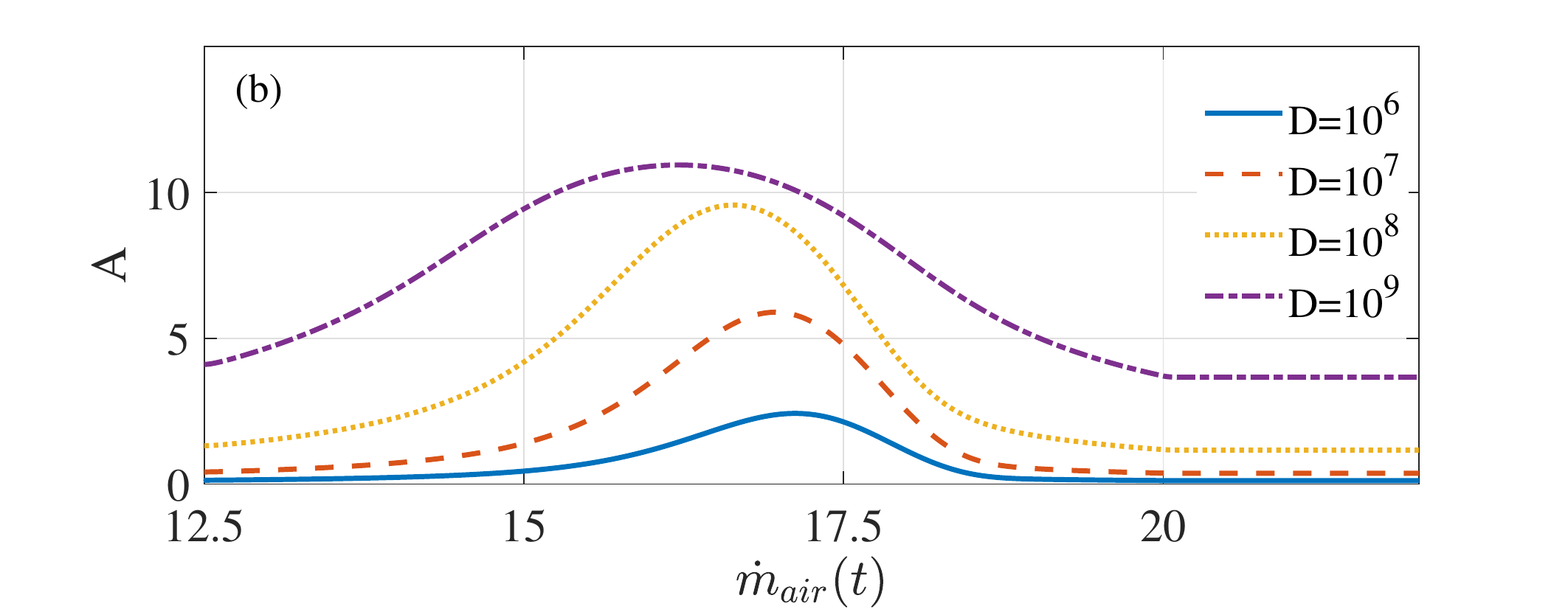}	
	\caption{The influence of different correlation time $\tau$ and intensity $D$ of noise on bifurcation dodge. (a) Different $\tau$ under the same $R=10$ and $D=8.75\times10^6$; (b) Different $D$ under the same $R=10$ and $\tau=0.001$. Other parameters are $\beta_1=-0.47$, $\beta_2=0.001$, and $\omega_{0}=100\times 2\pi$.}
	\label{fig3}
\end{figure}

\newpage
We have just analyzed that under the excitation of the colored noise, the effect of bifurcation dodge becomes better with the increase of the rate $R$. Next, the influence of the correlation time $\tau$ and the intensity $D$ of noise, which are important parameters to characterize the colored noise, on bifurcation dodge are studied. As shown in Figure \ref{fig3}(a), the larger the noise correlation time $\tau$ is, the lower the amplitude becomes, and the better the dodging effect of thermoacoustic instability is. Considering that white noise is only an ideal state, the correlation time of noise always exists. The existence of the noise correlation time $\tau$ is conducive to the avoidance of thermoacoustic instability in the thermoacoustic system, which shows that the actual control effect is better than that of the idealized mathematical model when the same $R$ is selected. Figure \ref{fig3}(b) shows that the amplitude of the thermoacoustic system increases due to the noise intensity $D$, which intensifies the pressure oscillation. Therefore, it is necessary to pay attention to control the noise intensity in a real thermoacoustic system.

\begin{figure}[htbp]
	\centering
	\includegraphics[width=8cm]{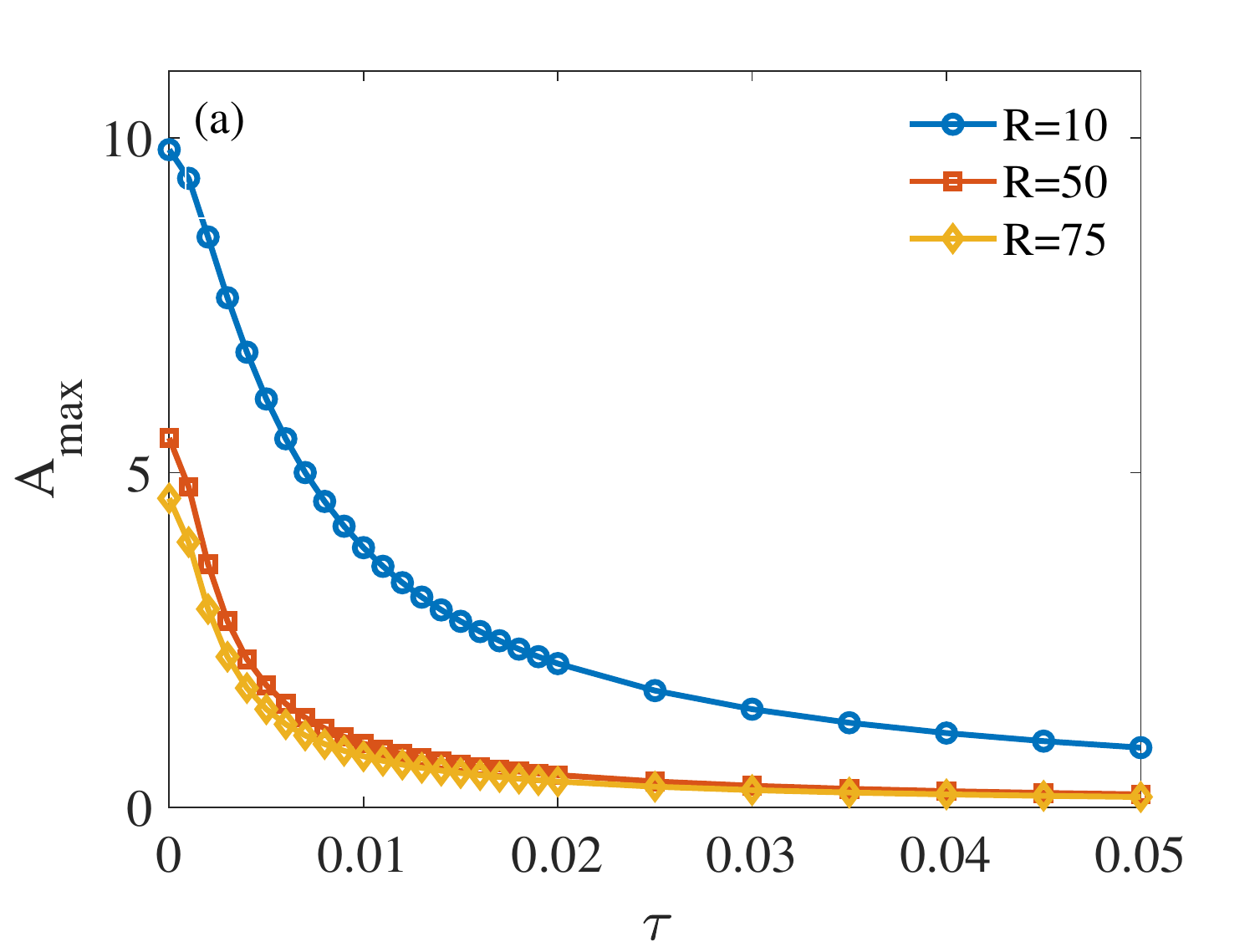}	
	\includegraphics[width=8cm]{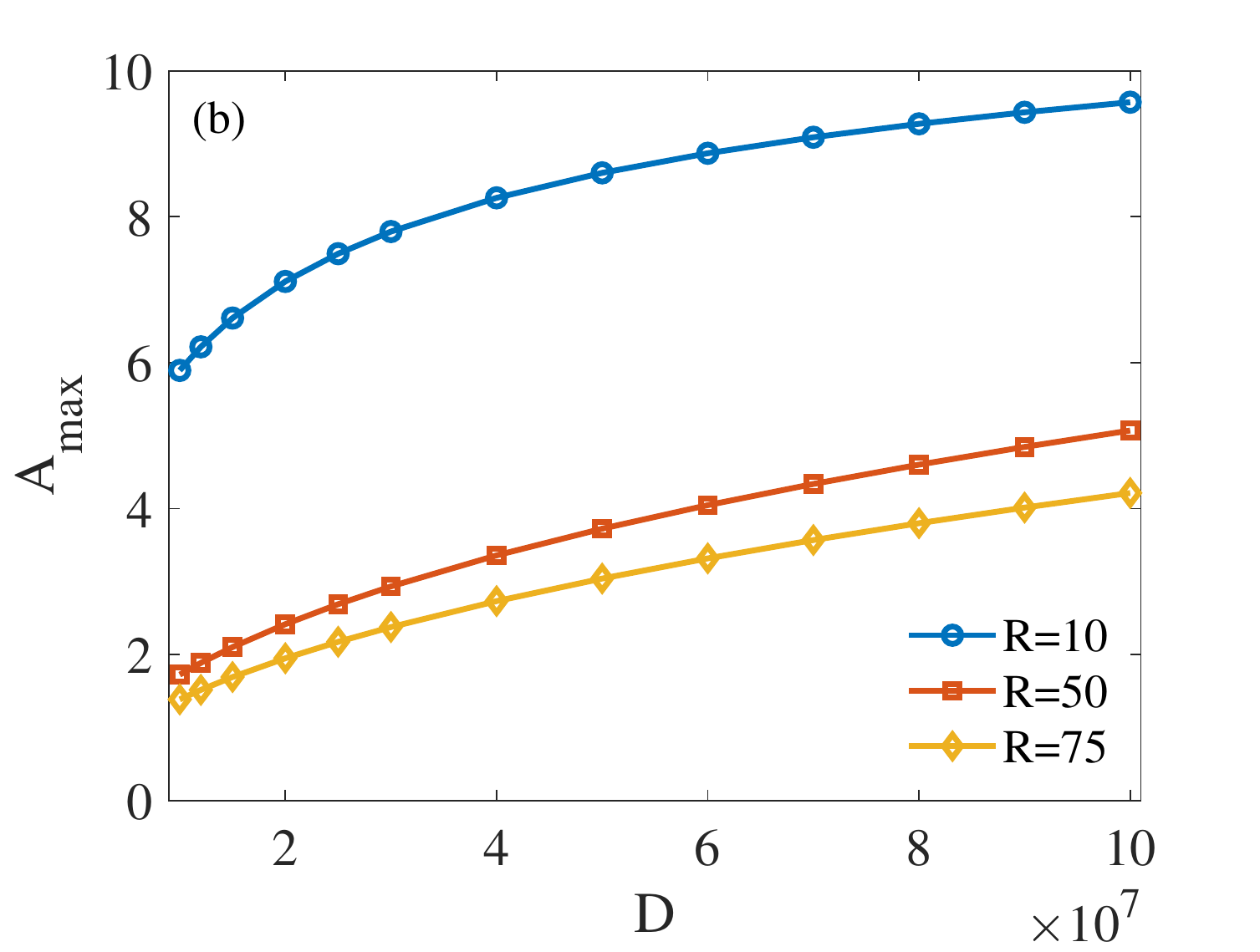}	
	\caption{Variation of the maximum amplitude $A_{max}$ with respect to (a) noise correlation time  $\tau$, and (b) noise intensity $D$ for different rates $R$. Other parameters are $\beta_1=-0.47$, $\beta_2=0.001$, and $\omega_{0}=100\times 2\pi$.}
	\label{fig4}
\end{figure}

Furthermore, we try to find a quantitative relationship between the amplitude of the system and the parameters  characterizing the noise when the rate $R$ is fixed, so as to better implement the control scheme. We consider the function of the maximum amplitude $A_{max}$ with respect to $\tau$ and $D$. It can be inferred from Figure \ref{fig4} that the maximum amplitude decreases rapidly with the increase of $\tau$, and then tends to be flat. With the increase of $D$, the rising speed of the maximum amplitude is also steep first and then slow.

\begin{figure}[htbp]
	\centering
	\includegraphics[width=8cm]{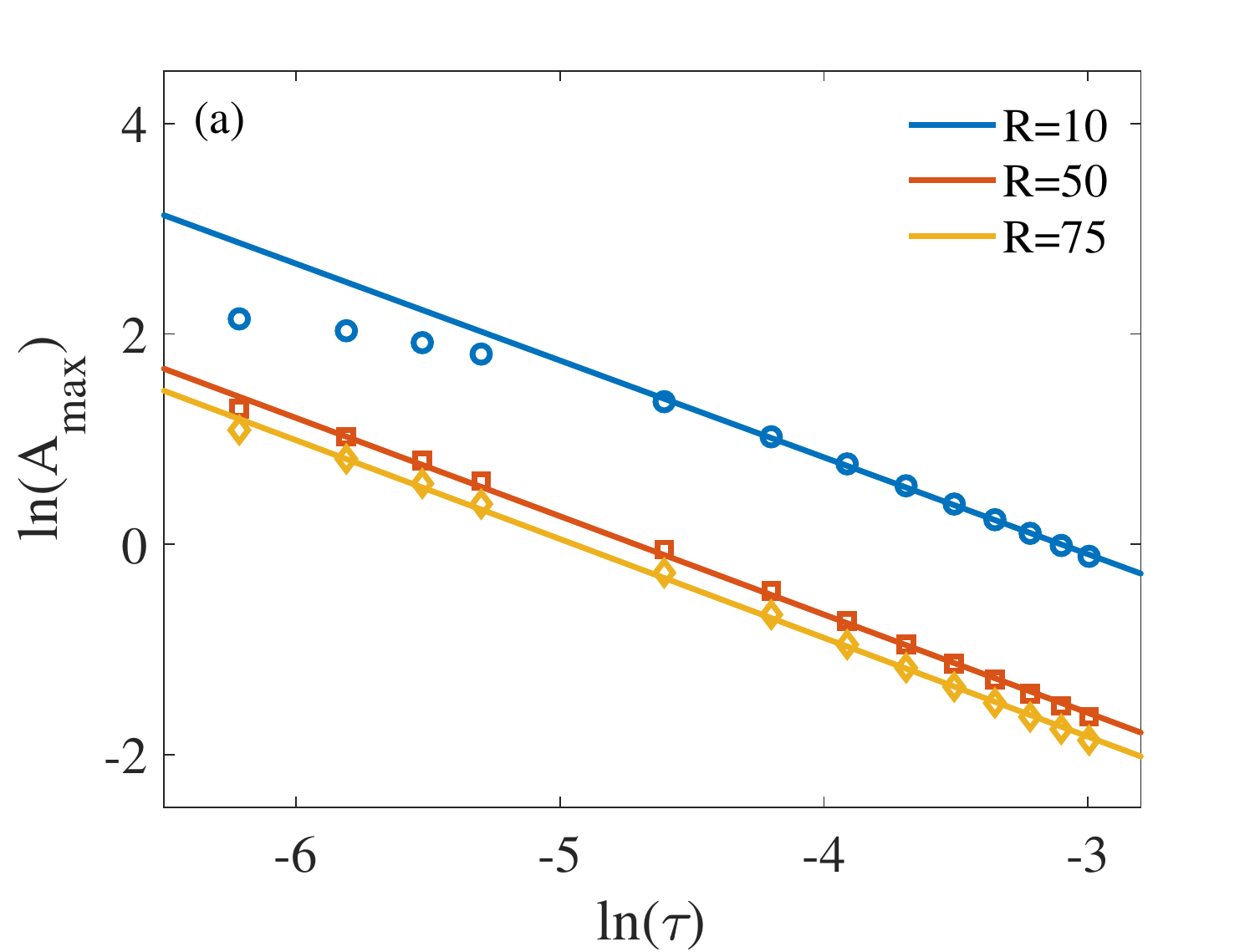}	
	\includegraphics[width=8cm]{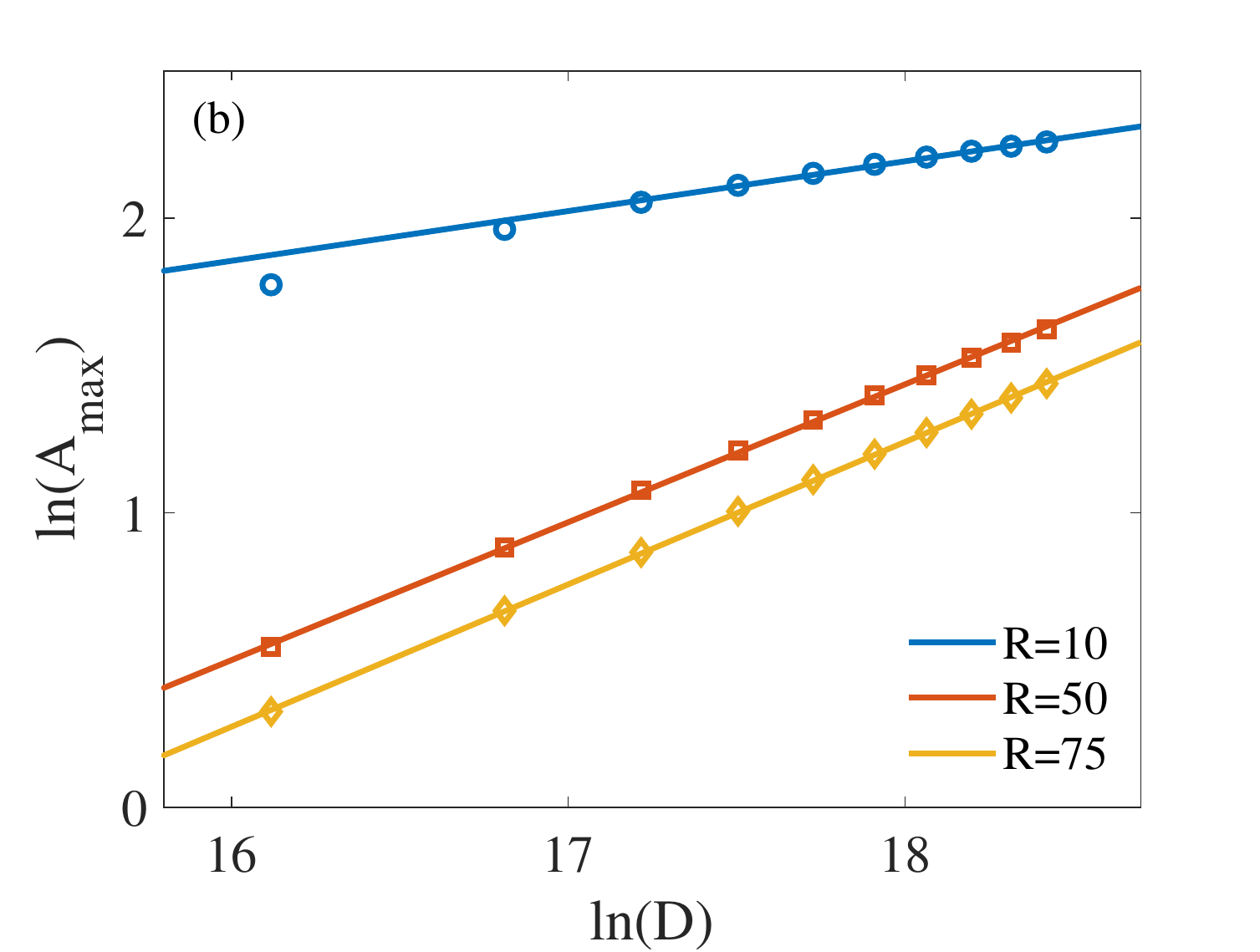}	
	\caption{(a) The log-log plot of the Eq. (\ref{logp}), between the maximum amplitude $A_{max}$ and the noise correlation time; (b) The log-log plot of the Eq. (\ref{logq}), between the maximum amplitude $A_{max}$ and the noise intensity $D$. The solid lines are fittings with linear functions.}
	\label{fig5}
\end{figure}

In order to determine more accurately the functional relationship, we draw the log-log plot of the maximum amplitude $A_{max}$ and both the noise correlation time $\tau$ and $D$ in Figure \ref{fig5}. We then clearly see that the relationship between the $ln(A_{max})$ and $ln(\tau)$ is a linear function. Let $ln(A_{max})=-p\cdot ln(\tau)+a$, where $p>0$ and $a$ is a constant. Then it can be deduced that
$$ ln(A_{max})=ln(\tau^{-p})+ln(e^a)=ln(e^a\cdot \tau^{-p}). $$
With logarithms removed from both sides of the equation, we get that $A_{max}$ and $\tau$ have the following power-law relationship 
\begin{equation}\label{logp}
	A_{max}(\tau) \sim \tau^{-p},
\end{equation}
where $p>0$ is the algebraic exponent. Similarly, we can obtain the power-law relationship between the maximum amplitude $A_{max}$ and the noise intensity $D$
\begin{equation}\label{logq}
A_{max}(D) \sim D^{q},
\end{equation}
where $q>0$ is the algebraic exponent. Here we notice that when the correlation time of noise $\tau$ is very small, i.e., the first few points in Figure \ref{fig5}(a), the power-law relationship is not well expressed. In this case, the colored noise is close to the white noise, and the randomness is relatively strong, so it is easy to destroy the power-law relationship. However, when $R$ is large, the rate becomes the major cause of the system qualitative change, which weakens the destructive noise impact caused by a small correlation time, as shown by the power-law relationship. But the power-law relationship is still relatively good when $R$ is large and the correlation time is small. 

\begin{figure}[htbp]
	\centering
	\includegraphics[width=12cm]{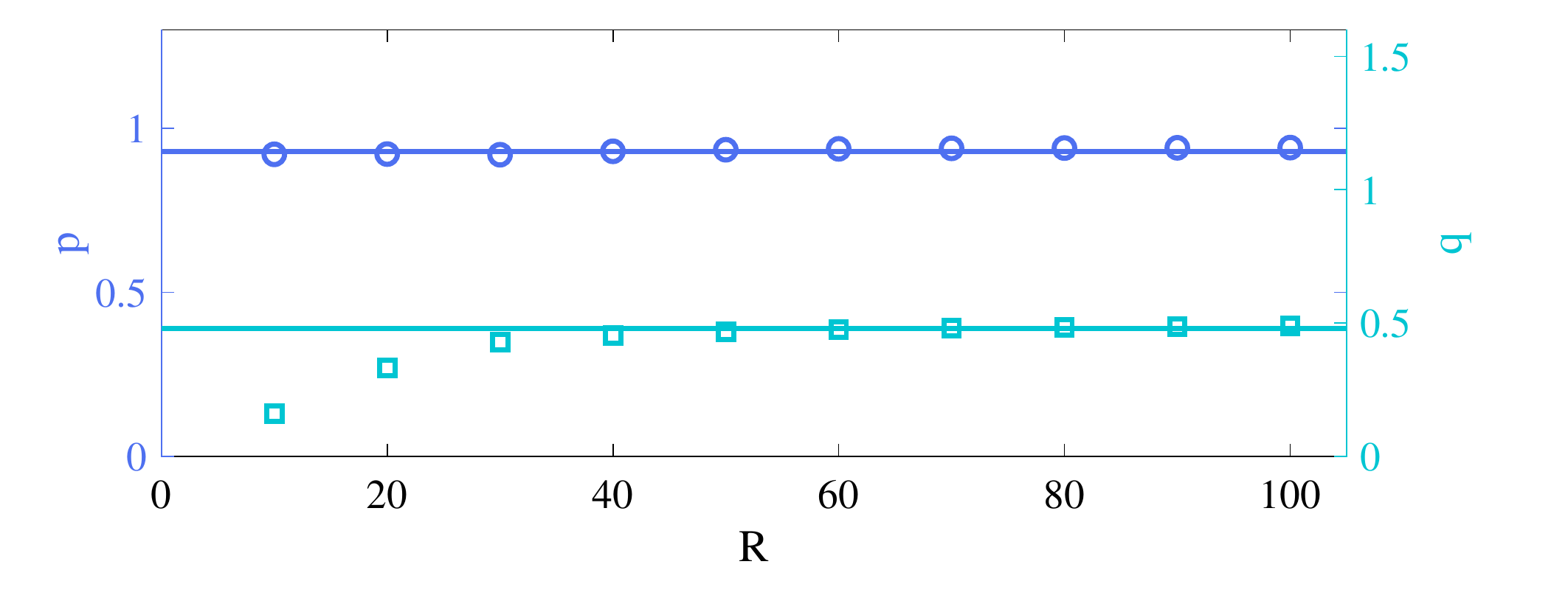}	
	\caption{Variation of the power exponents $p$ and $q$ of the Eqs. (\ref{logp}) and (\ref{logq}) calculated numerically for different parameter change rate $R$.}
	\label{fig6}
\end{figure}

For the sake of getting a more detailed relationship, we can estimate the values of the power-law exponents $p$ and $q$ under different rates $R$ by data fitting. We infer from the Figure \ref{fig6} that the power-law exponent $p$ of the noise correlation time remains virtually unchanged in the process of $R$ change. That is to say, $p$ is a global constant and the rate of decrease is constant for all rates $R$. In addition, the power law exponent $q$ of noise intensity is also stable at a constant value when $R>30$, but it decreases with the decrease of $R$ when the ramp rate is small. When the parameters change a little faster, the change speed of the maximum amplitude $A_{max}$ with the noise intensity $D$ is no longer affected by the rate $R$. This power-law relationship quantifies the influence of the correlation time of the noise and the noise intensity on the larger amplitudes in the thermoacoustic system, which is of highest significance for the control strategy to dodge the thermoacoustic instability.

\section{Probability of the rate-dependent bifurcation dodge}
\label{sec4}
For a practical thermoacoustic system, the threshold of thermoacoustic instability is usually defined according to its actual utility. So in this section we present calculations of the probability of a successful bifurcation dodge for a given threshold in the thermoacoustic system. 

\begin{figure}[htbp]
	\centering
	\includegraphics[width=13cm]{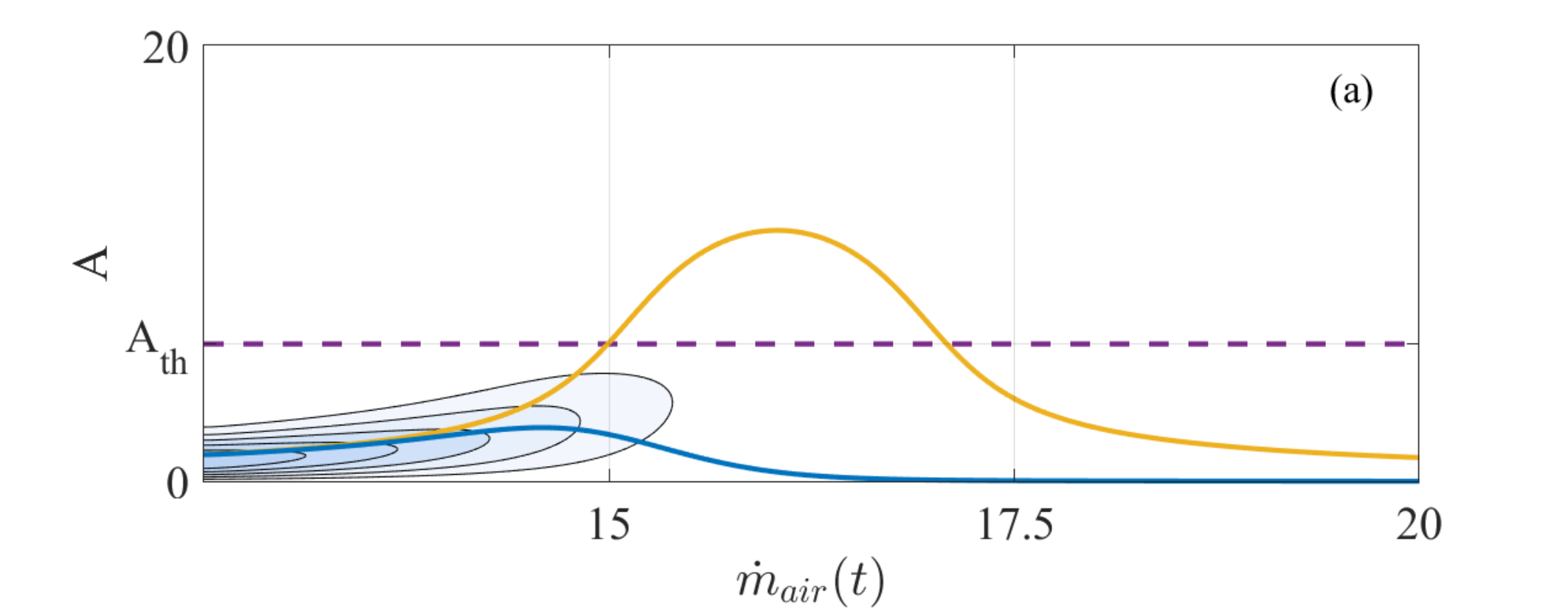}
	\includegraphics[width=13cm]{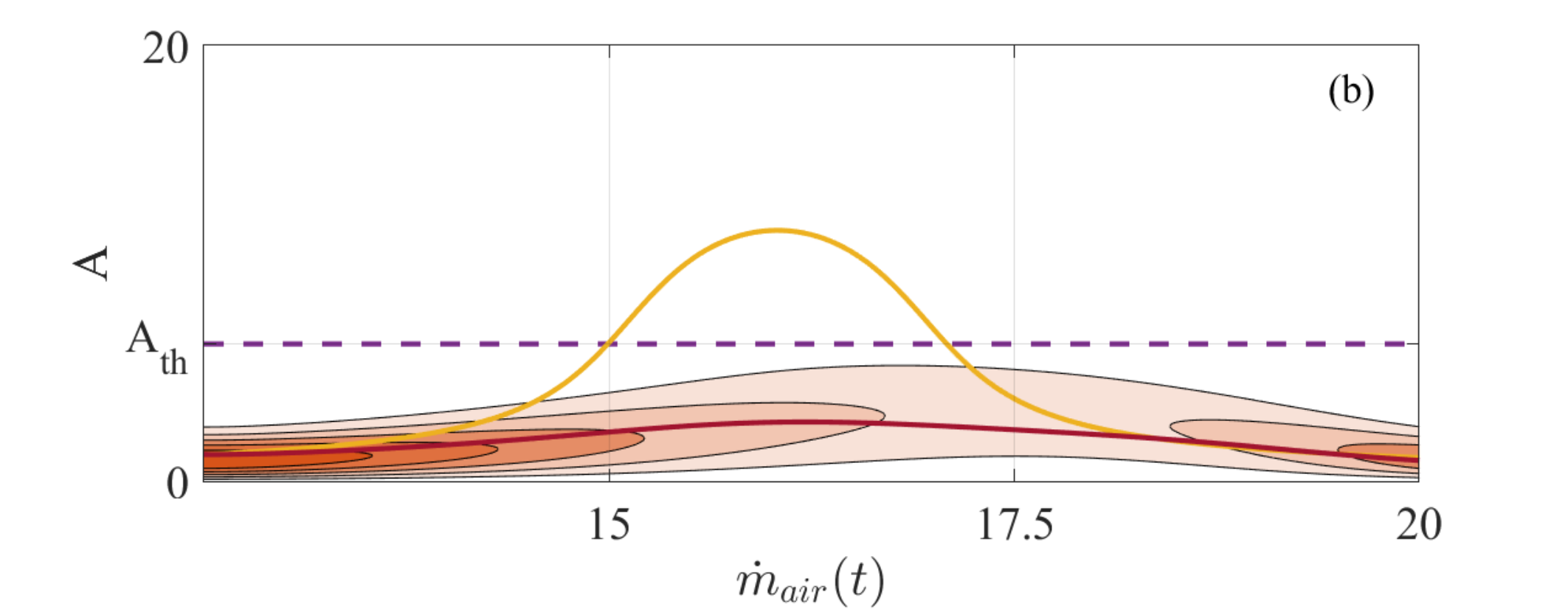}
	\includegraphics[width=13cm]{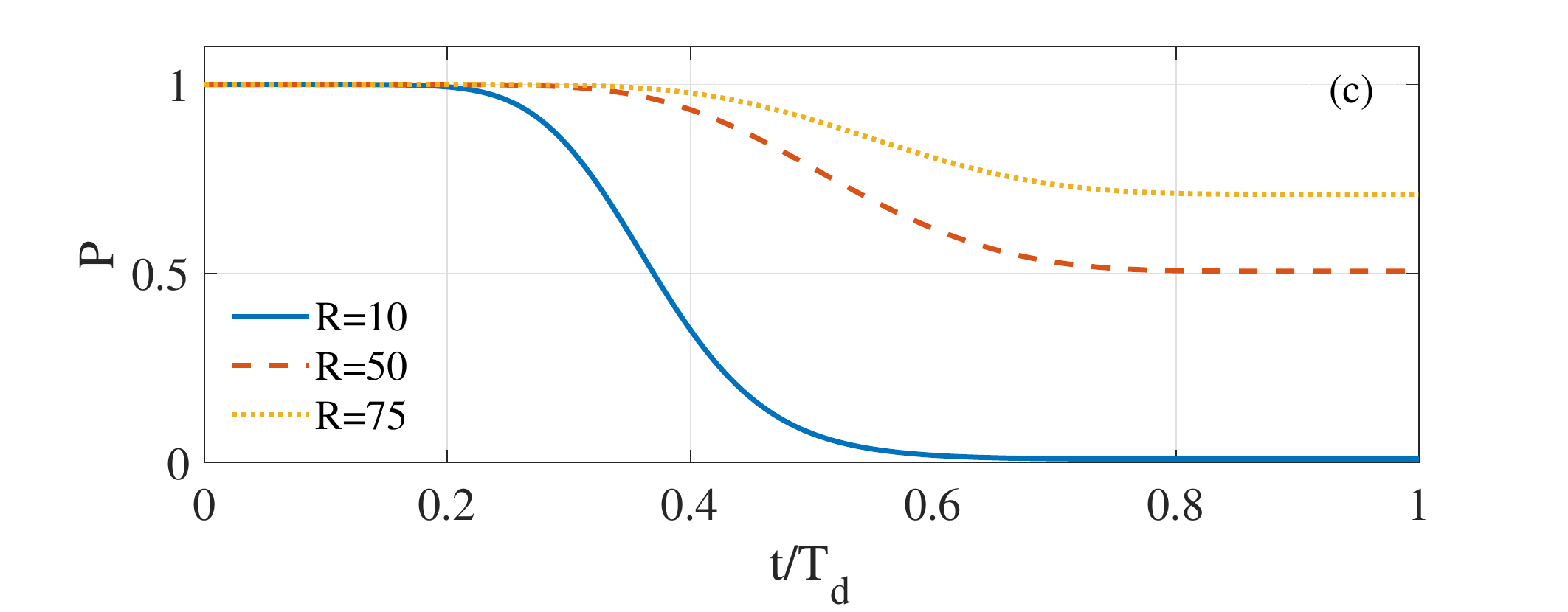}	
	\caption{The influence of different rates $R$ of the time-varying parameter on the transient dynamical behavior of the thermoacoustic system with a given threshold $A_{th}$. The yellow lines indicate the quasi-steady deterministic results as a reference. Dark blue and dark red lines represent the mean value of the amplitude $A$ under different $R$. (a) $R=10$; (b) $R=75$. (c) The probability of successful bifurcation dodge with different rates $R$ under the excitation of additive colored noise. For comparison, the abscissa is divided by the duration time of parameter change $T_d$ to make it unified in the same time scale. Other parameters are $\beta_1=-0.47$, $\beta_2=0.001$, $\omega_{0}=100\times 2\pi$, $D=8.75\times10^6$, and $\tau=0.001$.}
	\label{fig7}
\end{figure}

In the thermoacoustic system excited by additive colored noise, we consider that the threshold value of the thermoacoustic instability $A_{th}$ is between the maximum of the amplitude of the stationary state $A_{max}$ and the minimum value $A_{min}$. In this work, we consider the threshold $A_{th}=(A_{max}+A_{min})/2$ as an example to discuss the probability of rate-dependent bifurcation dodge. In fact, if the duration time of parameter change is $T_d$, the probability of a bifurcation dodge is the probability of the amplitude $A$ in the threshold range $[0,A_{th}]$ before time $T_d$. Thus, we can use the following formula to calculate the probability of  bifurcation dodge $P$ by the FPK equation (\ref{FPK})
\begin{equation}
P(T_d)=\int_0^{A_{th}} P(A,T_d) dA.
\end{equation}
Figure \ref{fig7} shows the influence of different rates $R$ of the time-varying parameter on the transient dynamical behavior of the thermoacoustic system with a given threshold $A_{th}$. Figures \ref{fig7}(a) and (b) are the results corresponding to Figures \ref{fig2}(a) and (b) after the threshold value $A_{th}$ is being set. It can be seen that when $R= 10$ (see Figure \ref{fig7}(a)), most of the amplitudes beyond $\dot m_{air}(t)=16$ exceed the threshold range, and the probability density is absorbed by the threshold boundary. However, when the rate is large (see Figure \ref{fig7}(b)), some of the amplitudes remain in the threshold range, which makes the probability of successfully dodging to reach $70\%$. Figure \ref{fig7}(c) shows the probability of successful dodging at different rates $R$. For the abscissa, we divide the time by the duration time of parameter change $T_d$, unifying them in the same time scale. It is not difficult for us to understand that the larger $R$ is, the higher the probability of successfully dodging thermoacoustic instability is. 

\begin{figure}[htbp]
	\centering
	\includegraphics[width=13cm]{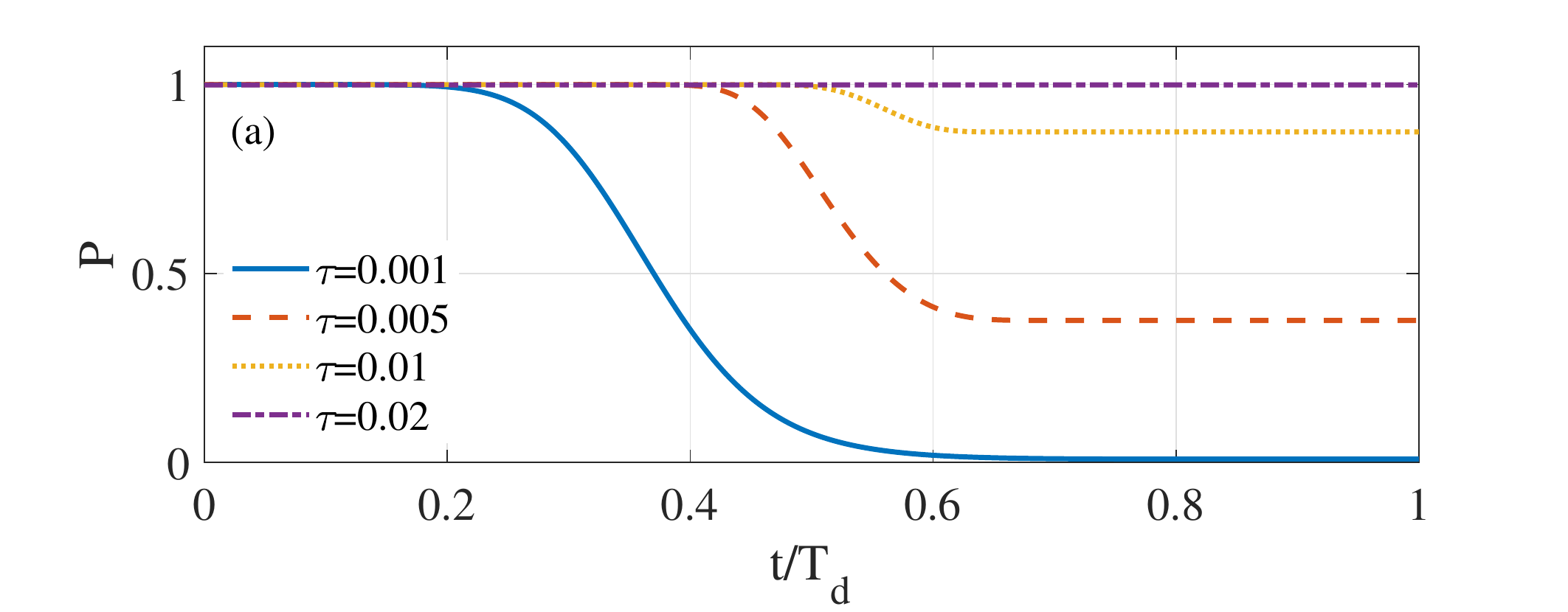}	
	\includegraphics[width=13cm]{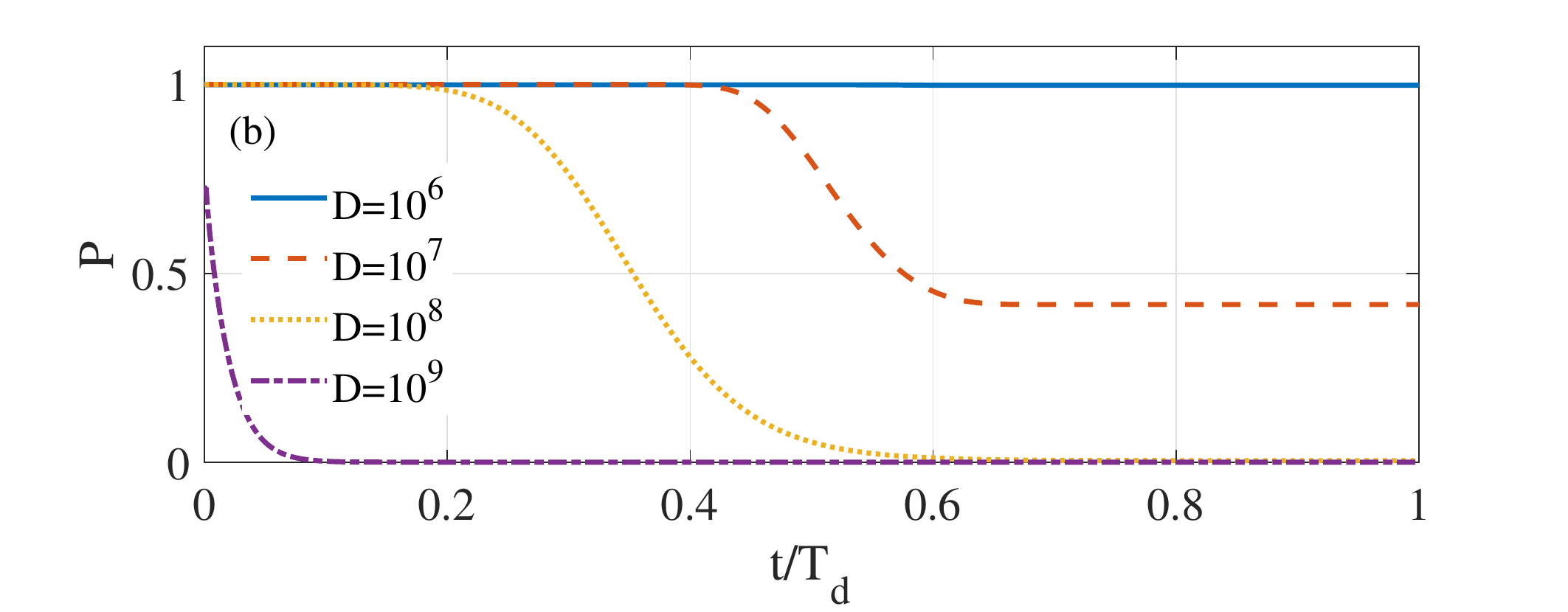}
	\caption{The probability of a successful bifurcation dodging under the excitation of additive colored noise with (a) different noise correlation time $\tau$, and (b) different noise intensity $D$. Other parameters are $\beta_1=-0.47$, $\beta_2=0.001$, $\omega_{0}=100\times 2\pi$, and $R=10$.}
	\label{fig8}
\end{figure}

Next, we consider the probability of the thermoacoustic system to successfully dodge the thermoacoustic instability threshold $A_{th}$ excited by additive colored noise under the condition of a fixed rate $R$ with different noise correlation time $\tau$ and noise density $D$. Figure \ref{fig8}(a) shows that the larger $\tau$ is, higher the probability of successfully dodging the high amplitude is. When the correlation time is very small, with the increase of time, it is almost impossible for the system to avoid the threshold of thermoacoustic instability. However, when $\tau$ increases to a certain value, the system can quench the thermoacoustic instability with a probability close to 1. This shows that the noise correlation time $\tau$ is a favorable factor to dodge the thermoacoustic instability. On the contrary, the noise intensity $D$ is not conducive to avoiding thermoacoustic instability. The higher $D$, the lower the probability of successfully avoiding thermoacoustic instability, as shown in Figure \ref{fig8}(b). 

\begin{figure}[htbp]
	\centering
	\includegraphics[width=8cm]{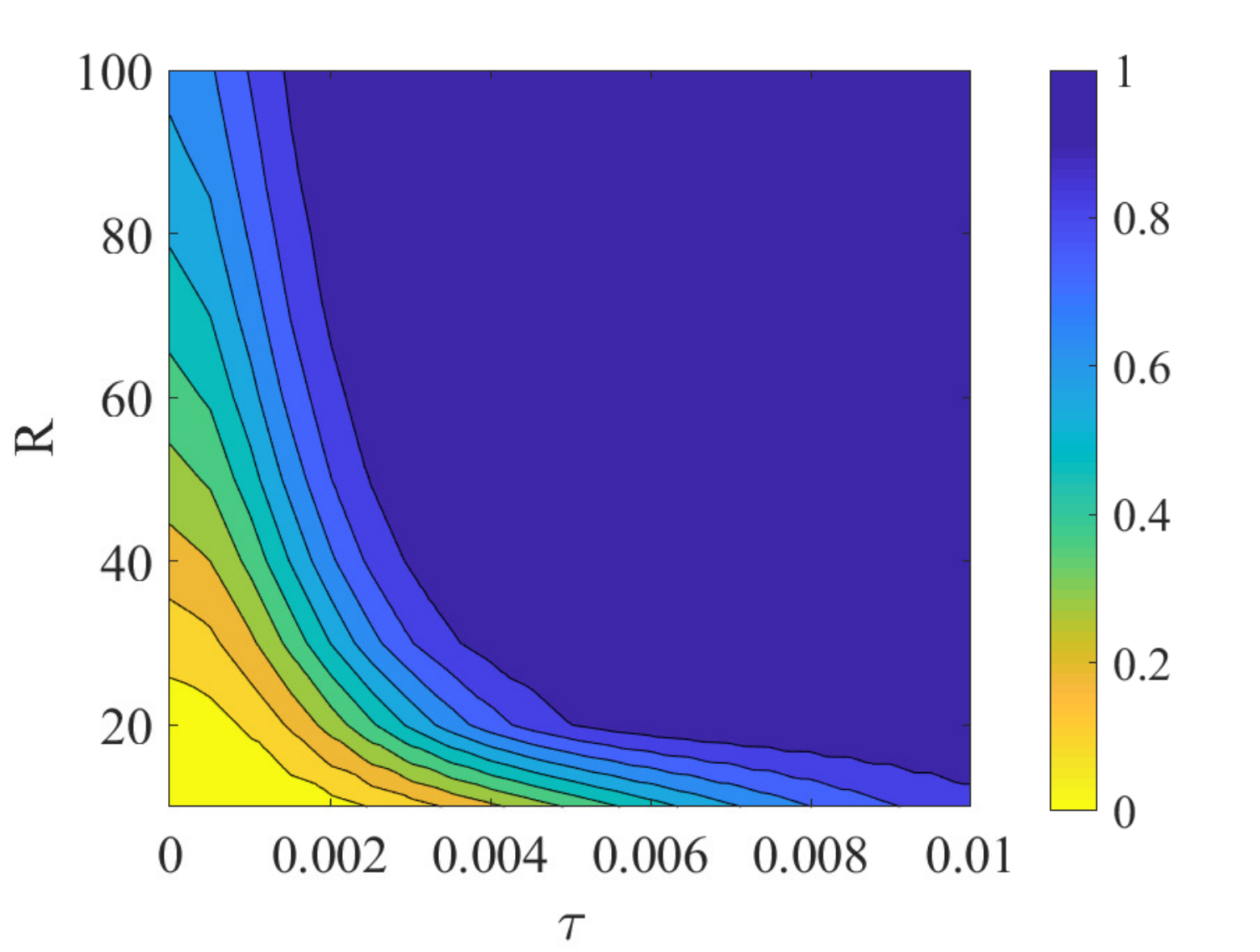}
	\includegraphics[width=8cm]{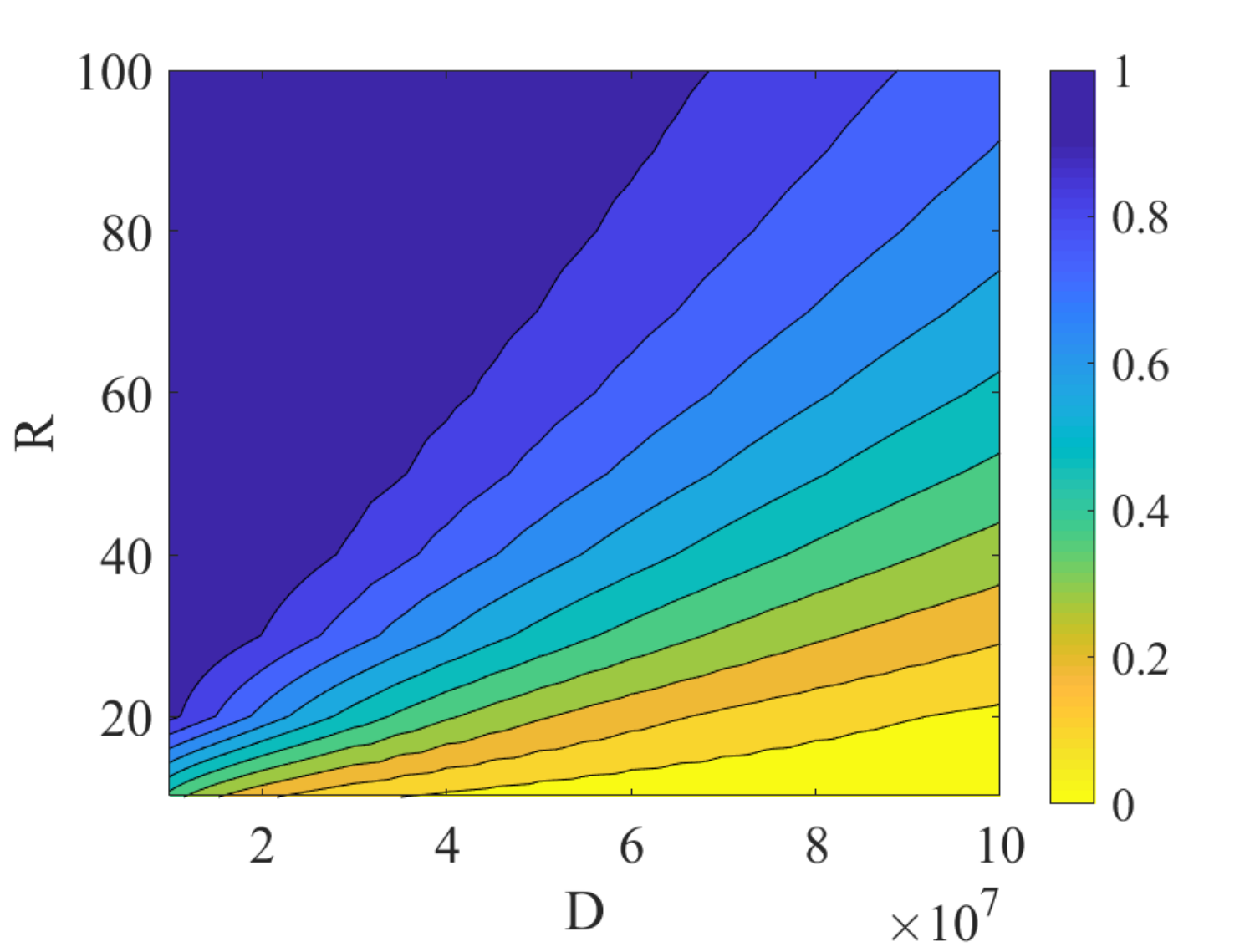}	
	\caption{The probability of the thermoacoustic system to successfully dodge the thermoacoustic instability when combined with the two factors under the excitation of additive colored noise. (a) The noise correlation time $\tau$ versus the ramp rate $R$; (b) the noise intensity $D$ versus the ramp rate $R$.}
	\label{fig9}
\end{figure}

When the rate $R$ and $\tau$ vary at the same time, the probability of a bifurcation dodging the instability is discussed in Figure \ref{fig9}(a). This gives that moderately accelerating $R$ and increasing $\tau$ can make the probability of the thermoacoustic system dodging the high amplitude to reach more than 90\%, so as to avoid the occurrence of thermoacoustic instability. Combined with $R$, we find that when $R$ is larger and $D$ is smaller, it is helpful to avoid thermoacoustic instability, as shown in Figure \ref{fig9}(b). When $D$ is relatively small or large, the parameter change rate has little effect on avoiding the thermoacoustic instability. That is to say, when the thermoacoustic system is affected by strong noise, the system  almost inevitably produces thermoacoustic instability. At this time, it is impossible to keep the system in a safe state by changing the rate of parameters. Hence, the noise intensity must be controlled first.

\section{Conclusions}
In this work, the avoidance of the undesirable state of a thermoacoustic system excited by colored noise is studied. Here we argue that there is a strong reason for assuming that the changing rate of the parameters affects the dodging of thermoacoustic instability, and colored noise is the most suitable form to describe the random forces in such rate-dependent system. We take into account the stochastic averaging method to overcome the non-Markov property of colored noise and to simplify the mathematical model. By means of the transient dynamical behavior shown by the evolution of the probability density, we find that the changing rate of parameter is conducive to dodge the thermoacoustic instability. Besides, we uncover that the increase of the noise correlation time makes the dodging effect better, but an increase of the noise intensity brings an adverse effect. Furthermore, the functional relationship between the rate of parameters, the correlation time of the noise, the noise intensity, and the maximum amplitude of the system are quantified in this work. The power-law relationship obtained is of utmost importance for understanding the internal mechanism of thermoacoustic systems and controlling thermoacoustic instability. Finally, the probability of successfully dodging the thermoacoustic instability under various conditions is calculated to validate the effectiveness of the control.

Our research is crucial to ensure a safe operation of thermoacoustic engines, because real nonlinear thermoacoustic systems are non-autonomous, the parameters do change with time and the correlation time of the noise can not be ignored. Considering that rocket engine, gas turbine and aeroengine are the main thermoacoustic systems, the cost of full-scale experiments is huge. Our theoretical research is helpful to save human and material resources. The rate of parameters, correlation time of the noise and noise intensity are used together to better regulate the thermoacoustic instability, which has a guiding role in the design of the engines and the control of the combustion process, and it is also of significance for the control of other systems with similar bifurcation properties.

\section*{Acknowledgments}
This paper was supported by the National Natural Science Foundation of China under Grant No.11772255, the Fundamental Research Funds for the Central Universities,  the Research Funds for Interdisciplinary Subject of Northwestern Polytechnical University, the Shaanxi Project for Distinguished Young Scholars,and Shaanxi Provincial Key R\&D Program 2020KW-013 and 2019TD-010. 

\section*{Conflict of interest}
The authors declare that they have no conflict of interest.

\bibliography{dodge_ref}
\bibliographystyle{alpha}

\end{document}